\documentclass[twocolumn,superscriptaddress]{revtex4-2}
\usepackage[T1]{fontenc}
\usepackage{graphicx}
\usepackage{amsmath}
\usepackage{hyperref}
\usepackage{caption}
\usepackage{subcaption}
\usepackage[dvipsnames]{xcolor}
\usepackage{booktabs}
\usepackage[normalem]{ulem}
\usepackage{enumitem}
\usepackage{makecell}
\usepackage{array}

\begin{document}

\title{Neural RHEED alignment with limited training data during CdTe MBE growth}
\author{Bartłomiej Turowski}
\author{Jakub J. Meixner}
\author{Róża Dziewiątkowska}

\affiliation{International Research Centre MagTop, Institute of
   Physics, Polish Academy of Sciences,\\ Aleja Lotnik\'ow 32/46,
   PL-02668 Warsaw, Poland}

\author{Wojciech Zaleszczyk}

\affiliation{International Research Centre MagTop, Institute of
   Physics, Polish Academy of Sciences,\\ Aleja Lotnik\'ow 32/46,
   PL-02668 Warsaw, Poland}
   
\affiliation{Institute of Physics, Polish Academy of Sciences,\\ Aleja Lotnik\'ow 32/46, PL-02668 Warsaw, Poland}

\author{Tomasz Wojciechowski}

\affiliation{International Research Centre MagTop, Institute of
   Physics, Polish Academy of Sciences,\\ Aleja Lotnik\'ow 32/46,
   PL-02668 Warsaw, Poland}

\author{Valentine V. Volobuev}
\affiliation{International Research Centre MagTop, Institute of
   Physics, Polish Academy of Sciences,\\ Aleja Lotnik\'ow 32/46,
   PL-02668 Warsaw, Poland}
\affiliation{National Technical University "KhPI"{},\\ Kyrpychova Str. 2, 61002 Kharkiv, Ukraine}

\author{Marcin M. Wysokiński}
\email{wysokinski@magtop.ifpan.edu.pl}
\author{Tomasz Wojtowicz}
\affiliation{International Research Centre MagTop, Institute of
   Physics, Polish Academy of Sciences,\\ Aleja Lotnik\'ow 32/46,
   PL-02668 Warsaw, Poland}

\date{\today}

\begin{abstract}
We introduce a data‑efficient neural‑vision assisted method to automate crystallographic alignment during molecular beam epitaxy (MBE) growth. Trained on reflection high-energy electron diffraction (RHEED) patterns from only 15 CdTe structures, our model - enabled by physics-aware postprocessing - reliably infers crystallographic directions, replacing manual frame‑by‑frame inspection. To this end, we design, test, and critically compare neural‑network architectures based on 2D and 3D ResNet configurations, both with and without postprocessing that leverages the physical constraints of RHEED image acquisition. Our work delivers (i) a fully trained neural system ready for closed-loop deployment in future CdTe growth experiments and (ii) a generalizable pipeline for new materials where access to diverse RHEED datasets is limited.
More broadly, this study represents a step toward AI‑driven MBE growth and demonstrates the potential of machine‑learning‑assisted automation in thin‑film synthesis.
\end{abstract}
\maketitle

\begin{figure}[h]
    \includegraphics[]{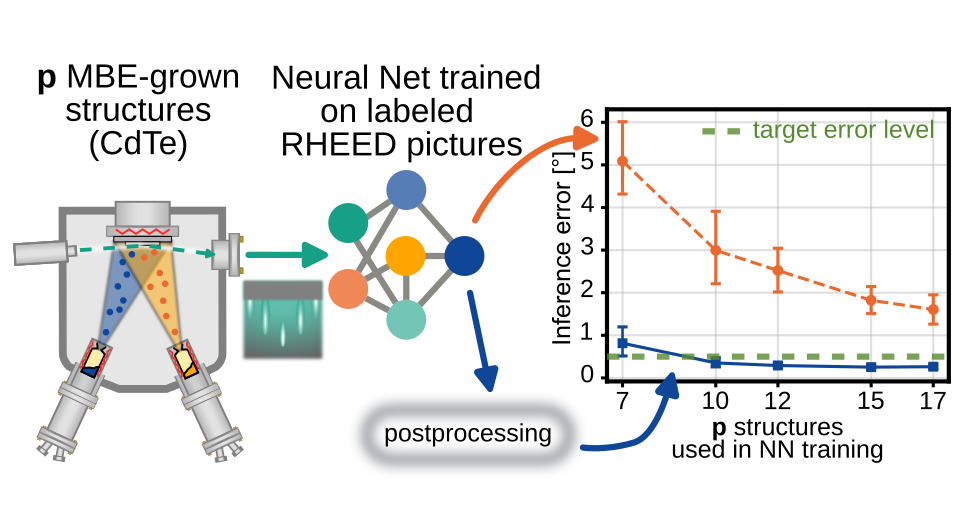}
\end{figure}
Abstract figure

\section{Introduction}
\label{sec:intro}
Molecular Beam Epitaxy (MBE) is a well‑established technique for producing high‑quality epitaxial thin films of semiconductors and related materials \cite{VVV2017, kazakov2025}. A critical early decision in any MBE process is the choice of substrate. Cadmium telluride (CdTe) buffer layers are widely used as epitaxial templates in both fundamental research \cite{Polaczynski_2024} and industrial fabrication of HgCdTe infrared detector structures \cite{Carmody2012, Pan2022}, with ML methods recently entering also the detector-design stage \cite{Bansal_2024}. The quality of the CdTe substrate directly determines the achievable structural and electronic properties of the target material. Although CdTe growth is considered mature, further optimization of growth parameters remains essential for improving reproducibility and device performance.

Traditionally, such optimization is performed manually by experienced MBE operators who supervise multiple growth cycles. This process is time‑ and resource‑intensive, and its reproducibility is affected by gradual drifts in system conditions during normal operation. Moreover, a single buffer‑layer growth typically spans several hours, making continuous human supervision impractical. These challenges motivate the development of automated monitoring tools that can support or partially replace manual oversight.

Reflection high‑energy electron diffraction (RHEED) is the main in‑situ technique used to assess surface structure and growth quality during MBE\cite{ChaoShen_2024}. The shallow incidence angle (1-5$^\circ$) and high electron energy (10-30 keV) provide strong surface sensitivity, producing characteristic patterns whose shape, intensity, and symmetry encode information about surface flatness, reconstruction, domain structure, and growth mode. The quantitative interpretation of RHEED patterns and intensity oscillations is supported by dedicated simulation codes \cite{Daniluk_2025, Daniluk_2026}. During growth, the substrate is continuously rotated to ensure uniform deposition.

Scientists are typically interested in only a few key azimuthal orientations - those aligned with crystallographic directions - because these produce symmetric, easily interpretable patterns. For this reason, full rotations are often recorded as short videos for later manual inspection. However, locating the key crystallographic directions within these recordings is time-consuming and prone to human error, leading to non‑negligible angular uncertainty. These limitations highlight both the inefficiency of current practice and the opportunity for automated analysis.

Deep convolutional neural networks have become a standard tool for the automated analysis of scientific images, including diffraction-based materials characterization \cite{Kaufmann_2020, Xiong_2024, Vecsei_2019} and atomically resolved electron microscopy \cite{Ziatdinov_2017}. Recent years have seen growing interest in applying machine learning (ML) and computer vision to RHEED analysis and to crystal growth in general \cite{Petkovic_2025}, starting from early big-data statistical analyses of full RHEED video streams \cite{Vasudevan_2014, Provence_2020, Gemperline_2025}. Existing approaches typically fall into two categories. The first uses static images of known crystallographic directions to classify RHEED patterns \cite{Kwoen_2020, Kwoen_2022, Muetzel_2026}, analyze as‑grown structures \cite{Chin_2025}, or perform automated phase mapping \cite{Liang_2022} and quantitative image analysis \cite{Yoshinari_2026}. The second focuses on videos acquired at fixed azimuthal orientations to identify initial growth modes\cite{Chong_2024, kim_2023}, detect growth‑mode transitions \cite{Kaspar_2025}, quantify growth kinetics from high‑speed RHEED streams \cite{Guo_2025}, or enable real‑time feedback control \cite{Shen_2024}. However, these methods rely on manually pre‑selected key directions, leaving most azimuthal information unused. Approaches based on external triggering \cite{Kwoen_2020, Kwoen_2022} require per‑growth calibration and remain susceptible to mechanical inaccuracies.

Only a few works have explored rotating‑substrate RHEED videos, and these have focused on rotation‑error detection \cite{Anjum_2023} or substrate de‑oxidation monitoring \cite{walieh_2023}, which lately was pushed further with real-time feedback control \cite{Shen2024}. Very recently, neural reconstruction of azimuthal RHEED profiles acquired during wafer rotation has also been demonstrated \cite{walieh_2025}, aimed at visualizing the azimuthal information rather than at quantitatively extracting the crystallographic directions. To date, there is a notable lack of methods capable of automatically identifying crystallographic directions directly from rotating RHEED data, either during growth or in post‑growth analysis \cite{Yu2025}. Such automation would reduce dependence on expert supervision and facilitate more efficient use of RHEED data in both research and production environments and would close the gap between substrate de-oxidation \cite{walieh_2023, Shen2024} and automatic growth optimization that rely on static RHEED data of single crystallographic direction \cite{Chin_2025, Chong_2024, Kaspar_2025, Yu2025}.

In this work, we leverage 2D and 3D convolutional residual network (ResNet) neural networks to show that, even when trained on a limited number of data points coming from no more than 20 different structures, they can be practically useful for detecting crystallographic directions from RHEED images of CdTe grown by MBE. In particular, thanks to a proposed postprocessing step applied after neural network inference, we demonstrate that a simpler 2D ResNet trained on 15 structures can be as accurate as a complex 3D ResNet trained on all available structures (20) without this step. At the same time, the 2D approach outperforms the 3D model in training time efficiency by a factor of $\sim$20. From a more general perspective, by using the proposed pipeline, we demonstrate reliable automated identification of crystallographic directions from rotating‑substrate RHEED videos, providing a practical step toward artificial intelligence (AI)‑assisted and, eventually, autonomous MBE growth.

The article is structured as follows: Section~\ref{sec:frame} addresses the problem, followed by Sections~\ref{sec:data} and \ref{sec:hardware}, which detail data gathering, preprocessing steps, and the hardware used for neural network training and real-time inference. Sections~\ref{sec:methodology} and \ref{sec:results} outline the neural network methodology and present the results and discussion, respectively, while Section~\ref{sec:summary} provides a final summary.

\section{Framing the problem}
\label{sec:frame}
The goal of acquiring RHEED images during MBE growth is to continuously assess the quality of the film and, when necessary, adjust growth conditions or terminate the process to avoid wasting time and material. Crystallographic directions that align with crystallographic axes of the grown material provide the most informative geometry for evaluating growth quality, because the resulting diffraction patterns are highly symmetric and can be interpreted quickly by a trained observer.

An ideal long‑term solution would be a neural‑assisted system capable of detecting growth anomalies in real time and autonomously tuning MBE parameters to optimize the final film quality or to halt the growth when recovery is not possible. First steps in this direction have already been demonstrated for related thin-film techniques, including machine-learning-driven closed-loop optimization of epitaxial growth \cite{Ohkubo_2021, Wakabayashi_2019} and autonomous synthesis workflows \cite{Harris_2024, Zheng_2025}. To enable such a system, one must reliably collect RHEED images corresponding to the same crystallographic direction at different stages of the growth. This consistency is also essential for traceability and explainability, allowing a human expert to verify the neural system's decisions.

However, simply aligning the sample to a chosen crystallographic direction at the beginning of the growth is not sufficient. Over time, small mechanical drifts or changes in beam incidence can shift the relative orientation between the electron beam and the sample surface. As a result, RHEED images taken at different times may no longer correspond to the same direction.

Therefore, the challenge addressed in this work is to use a trained neural network to identify the crystallographic direction within each set of RHEED images acquired during a full 360$^\circ$ rotation. This enables the images to be correctly grouped across different growth times, compared consistently, and prepared for subsequent automated analysis.

A further difficulty, closely related to the problem framed above, arises from the limited amount of RHEED data typically available in an MBE laboratory. While obtaining a RHEED dataset containing 10–20 structures of a selected material system is achievable, assembling a well curated library of hundreds of structures is a time-consuming and labor-intensive task beyond the capacity of a typical MBE laboratory. Extending such a library to cover multiple material systems makes this practically unfeasible. Although neural networks are known to perform exceptionally well when supplied with large, diverse datasets, their performance falls rapidly when data is limited. In this work, we therefore show how to achieve reliable, real-time identification of crystallographic directions in RHEED images while training on data from only about 15 structures - demonstrating that meaningful performance is achievable even with limited data available.

\section{Data}
\label{sec:data}
A critical component of this work is the acquisition of high‑quality data. This naturally separates into two streams: collecting reliable RHEED images during the growth of high‑quality CdTe samples, and preprocessing these images into a form best suited for neural‑network ingestion. We begin by focusing on the former.

\subsection{Experimental data acquisition}
\label{sec:acquisition}
CdTe samples were grown using standard MBE procedures in an ultra-high-vacuum Veeco GENxplor\cite{GENxplor} system chamber dedicated to II–VI semiconductors. Effusion cells were loaded with elemental Cd, Zn, and Te (7N purity), and CdTe buffers were deposited on epi-ready GaAs(001) wafers with a $2^\circ$ off-cut toward the $\langle 110\rangle$ direction after oxide removal by hydrogen-plasma treatment at $200^\circ$C a separate chamber. A thin ZnTe nucleation layer (5–20 nm) was first deposited, followed by a $\sim 4\,\mu\mathrm{m}$ CdTe layer in the zinc-blende structure, with the substrate continuously rotated at $\geq 2$ revolutions per minute (RPM) and slowed to 1 RPM during RHEED video acquisition. Throughout the several-hour growth, 2-minute RHEED videos were automatically recorded every hour using the Molly 2000 software \cite{Molly2000} to trigger the RHEED shutter, while a custom screen-grab utility ensured consistent capture of the same screen region with fixed camera geometry. Videos were saved using the \texttt{avc1} codec at 18 frames per second (FPS), and experienced operators later identified the zero-angle ([011] direction), after which 1080 frames - corresponding to one full rotation - were extracted and labeled, yielding $0.33^\circ$ angular resolution. For training purposes, 151 single-rotation videos from growths of 20 CdTe buffers, 6--8 videos per structure, were used (cf. Table~S1).

Because the RHEED data were obtained via real‑time screen grabbing, there is a mismatch between the 60 Hz monitor refresh rate and the $\sim$18 FPS camera stream: $60/18 \approx 3.33$, causing some (around 5 to 10\%) frames to be duplicated or skipped. This can cause adjacent frames to appear identical despite having different crystallographic labels, slightly reducing the achievable accuracy of the neural‑network model. Consequently, the ground-truth labels themselves carry an uncertainty of at least one frame ($0.33^\circ$), stemming from the manual identification of the zero-angle direction and from the duplicated or skipped frames. The inference errors reported below should be interpreted relative to this label noise floor.

Additionally, there are systematic issues, shown in Figure~S1, arising from MBE system geometry and everyday use:

\begin{itemize}
\item the GENxplor system’s three-handle molyblock holder introduces a periodic obstruction every $120^\circ$;
\item and a broad shadow near the RHEED beam edge originates from material flakes accumulating on the heater area.
\end{itemize}

Most of these issues are mitigated during data preparation and in our preprocessing pipeline, but they remain an inherent limitation of the raw data. A list of all grown structures with the number of rotations and frames after data preparation is presented in Supporting Information\cite{supp_info} Table S1.

\subsection{RHEED images preprocessing}
\label{sec:preprocessing}
\begin{figure}[b]
    \centering
    \includegraphics[width=0.48\textwidth]{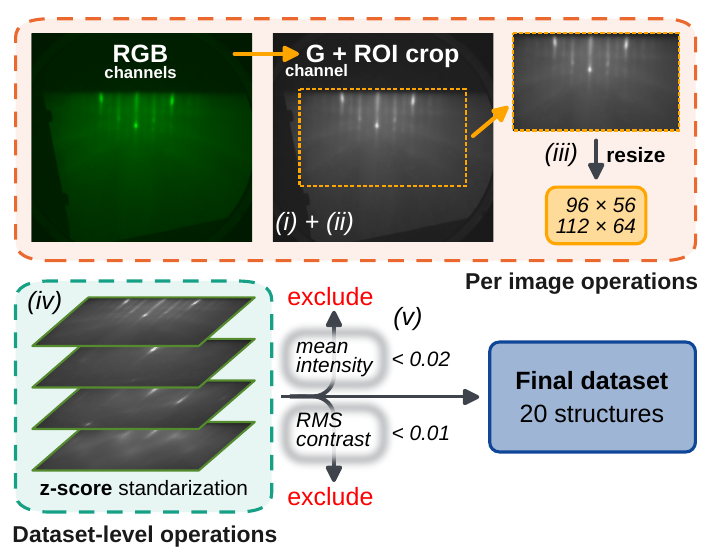}
    \caption{Schematic overview of the dataset‑preparation procedure, with items (i–v) corresponding to the steps detailed in Section~\ref{sec:preprocessing}. }
    \label{roi_area}
\end{figure}

\begin{figure}[b]
    \centering
    \includegraphics[width=0.48\textwidth]{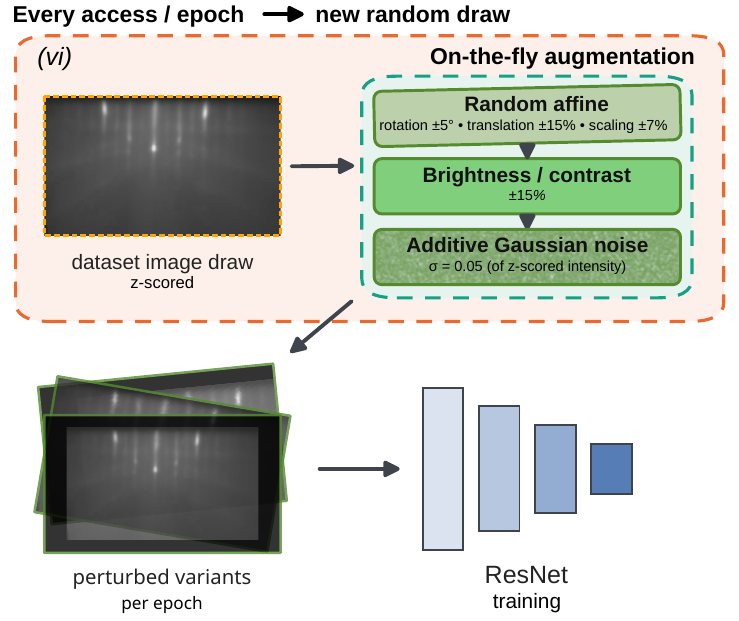}
    \caption{Schematic flow illustrating the on‑the‑fly replacement of each drawn training image by its augmented counterpart (the dataset size remains unchanged). The augmentation pipeline includes several transformations, with translation being the key variable dominating the ROI‑placement uncertainty.}
    \label{on-the-fly-augmentation}
\end{figure}

Raw RHEED frames were acquired at a resolution of \(1022\times 964\) in RGB uint8 pixel format. The final dataset was prepared using pipeline described below and shown schematically on Figure~\ref{roi_area}.
\begin{enumerate}[label=(\roman*), wide, labelwidth=!, labelindent=0pt]
    \item The RGB image data from screen capture is deterministically \textbf{converted to a single channel}, a process that does not cause information loss as the aggregated saturation of green color components in our dataset is approximately \(97\%\). Intensities were subsequently scaled to \([0,1]\). Decomposition of a single RHEED frame into three channels is shown in Figure~S2.
    \item A \textbf{fixed proportional crop} was then applied in image coordinates - top \(23\%\), left \(13\%\), right \(10\%\), bottom \(30\%\) - to remove peripheral artifacts while preserving the central diffraction content in the region of interest (ROI).

     \item The cropped \textbf{images were resized} to two target resolutions of \{\(96\times56\), \(112\times 64\)\} (width \(\times\) height) with approximately the same scaling ratio for both dimensions. The resizing procedure was done using bicubic interpolation to float32 format \cite{Keys1981}.

    \item After crop and resize, \textbf{dataset-level standardization} was performed on the grayscale channel: letting \(x\in[0,1]\) denote a pixel intensity, the standardized value was \(z=(x-\mu)/\sigma\), where \(\mu\) and \(\sigma\) are the empirical mean and standard deviation estimated across all retained pixels in the training split after preprocessing.
\item To suppress \textbf{unusable frames} prior to statistics, any frame with global mean intensity \(<0.02\) (under-exposed) or global standard deviation \(<0.01\) (insufficient RMS contrast) was \textbf{excluded}.
\item In order to improve neural network (NN) robustness in real time inference use, the \textbf{augmentation has been introduced} on top of the final data set created by applying steps (i-v) but before the training (cf. Figure~\ref{on-the-fly-augmentation}). To emulate the operator-dependent variability in how the ROI is placed over the RHEED video stream, each training image was replaced on-the-fly by a randomly perturbed variant - re-sampled independently at every access, so that across epochs the network is exposed to a continually varying version of each frame rather than the original frames, without enlarging the dataset. The perturbations combined small affine transformations (rotation $\pm 5^\circ$, translation $\pm15\%$, scaling $\pm7\%$) with brightness/contrast variation ($\pm15\%$) and additive Gaussian noise ($\delta$ = 0.05), applied directly on the standardized (z-scored) images without re-clipping their intensity range.

\end{enumerate}

The full pipeline was applied consistently to all images to ensure statistical comparability and reproducibility across the dataset.
\section{Hardware}
\label{sec:hardware}

Data acquisition is done on the RHEED computer supplied by k-Space Associates, Inc. kSA 400 software was used to display the data, but the acquisition method is independent of the software due to use of a custom screen-grab utility. The same computer was used to implement the inference in the real system.

\begin{table}[ht]
  \caption{Hardware (including operating system, processing units and memory) used in data acquisition, NN training and inference tests.}
  \label{tbl:hardware}
  \centering
  \renewcommand{\arraystretch}{1.4}
  \resizebox{\columnwidth}{!}{%
  \begin{tabular}{lll}
    \hline
    & Training  & \makecell[l]{Data acquisition \\ and inference}  \\
    \hline
    OS   & \makecell[l]{AlmaLinux 9.5 \\ (Teal Serval)} & Windows 10 Pro  \\
    CPU  & \makecell[l]{AMD Ryzen Threadripper \\ PRO 5955WX} & Intel Core i5-6500 \\
    RAM  & 128 GB  & 8.00 GB \\
    GPU & \makecell[l]{2x NVIDIA RTX A6000 \\ 2x (48 GB)} & \makecell[l]{AMD FirePro W4100 \\ 2 GB} \\
    \hline
  \end{tabular}}
\end{table}

Training was done on a single GPU while NN inference tests were run on CPU (cf. Table \ref{tbl:hardware}).
 
\section{Neural Network Methodology}
\label{sec:methodology}
In the present work, rather than designing a new neural‑network architecture for automated RHEED pattern recognition, we adopt established convolutional models - specifically, variants of the ResNet family \cite{He_2016}. Using off‑the‑shelf architectures enables us to leverage models that have been extensively validated across diverse image‑classification tasks \cite{Xie_2017, Liu_2022_ConvNeXt, Tan_2021_EfficientNetV2, Dosovitskiy_2021, Liu_2021_Swin, Touvron_2021_DeiT}, allowing us to focus on the goal outlined in Section \ref{sec:frame}. 
Conceptually, we compare two distinct approaches:
\begin{itemize}
\item \textbf{The two‑dimensional ResNet‑based approach}, combined with extensive postprocessing, which enables accurate inference even with limited training data. In our work, this serves as \textbf{the main architecture.}
\item \textbf{The three-dimensional ResNet-based approach}, which is well suited to the task but computationally heavier and reliant on larger datasets. In our setting, this serves as \textbf{the benchmark architecture.}
\end{itemize}
 
\subsection{2D ResNet approach}
\label{sec:resnet2d}
The most natural approach for computer‑vision‑assisted recognition of two‑dimensional images is to use a neural network that takes a 2D matrix of pixel intensities as input and returns a desired label as output - in our case, the angular deviation from a crystallographic direction. A well‑established architecture that fits this paradigm is the two‑dimensional convolutional ResNet. Below in steps we outline our approach, whose effectiveness heavily relies on post‑processing of the network’s raw predictions:
\begin{enumerate}[wide, labelwidth=!, labelindent=0pt]

    \item {\bf Dataset construction}
    
    We construct a data set $\{P_k,\alpha_k\}$, where $P_k$ denotes the $k$-th RHEED image and $\alpha_k$ is the corresponding label, defined as the sample's rotation angle relative to a chosen crystallographic direction. Since RHEED patterns recorded at azimuths differing by $180^\circ$ are very similar, all angle labels are folded modulo $180^\circ$, i.e. $\alpha_k \in [0^\circ, 180^\circ)$, even though each recorded video covers a full $360^\circ$ rotation. Consequently, the inference of the trained network is also restricted to this range. We have collected a labeled data set coming from 20 different CdTe structures. To evaluate generalization across structures, we repeatedly split the dataset by selecting $p$ structures for training and the remaining $20-p$ for the holdout set - the method descriptively called {\it repeated train-test split} approach, following established practice for avoiding data leakage in small datasets \cite{Wang_2020}. We explore splits ranging from $p=7$ to $p=17$. For simplicity, we shall use notation such as "12/8", meaning that 12 structures are used for training and 8 for holdout data sets. From each training set consisting of $p$ structures we are extracting 2 randomly selected structures as a validation data set, used exclusively for checkpoint selection, i.e. picking the epoch with the lowest validation mean absolute error (MAE).  While increasing the validation set to three structures reduces sensitivity to single-structure outliers, it incurs a non-negligible loss in effective training data. We therefore retained a two-structure validation set across all splits to maximize training efficiency.   
    
 To quantify performance on the holdout set (consisting of $20-p$ structures), 
we define $n$ as the total number of test samples. Each test sample corresponds 
to a single rotation of a given structure in the MBE chamber. The value of $n$ 
is determined by the number of rotations per structure, multiplied by the number 
of holdout structures, and by the fact that each holdout rotation is resampled 
10 times, i.e.
\[
    n = (\text{rotations per structure}) \times (20-p) \times 10.
\]
For overview see schema in Figure~\ref{dataset_division}.  To characterize the distribution of the results, we compute the mean inference error alongside $95\%$ bootstrap confidence intervals of the mean, estimated using the percentile method with 2000 resamples. The bootstrap procedure was applied to the mean errors of individual evaluation runs. Each run used an independent train/holdout split and a distinct random seed. This accounts for uncertainty stemming from data partitioning.

    \begin{figure}[h]
    \centering
    \includegraphics[width=0.5\textwidth]{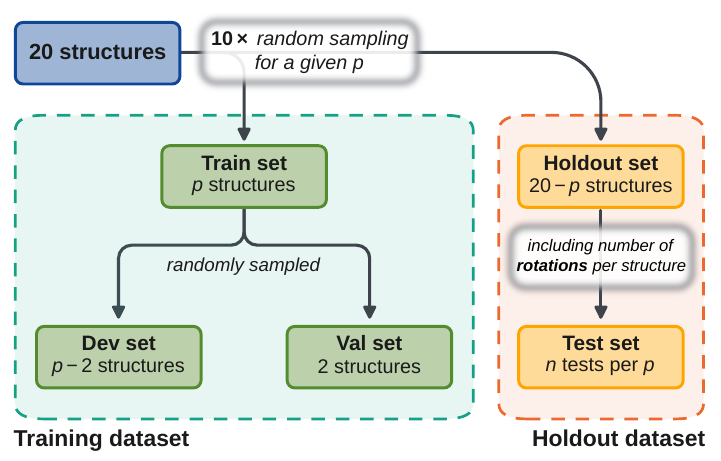}
    \caption{Visualization of the division of images from the initial 20 structures into development, validation, and holdout datasets. The parameter $n$ denotes the total volume of the testing dataset and reflects the number of rotations per structure, the number of structures sampled for the holdout set, and the fact that this sampling is repeated 10 times.}
    \label{dataset_division}
    \end{figure}
    
    \item {\bf Network training}
    
The 2D ResNet NN is trained to minimize the deviation of $NN(P_k)-\alpha_k$ over the entire dataset, where $NN(P_k)$ denotes the angle predicted by the network. Training is performed using the mean‑squared‑error (MSE) loss function. Treating the angle as a scalar regression target introduces an artificial discontinuity at the $0^\circ/180^\circ$ boundary, which the MSE loss penalizes heavily. We deliberately retained this simple formulation: the resulting wrap-around errors manifest as isolated outliers, which are subsequently removed by the postprocessing step.

We initialized the ResNet-50 backbone with weights pre-trained on ImageNet (pyTorch \texttt{IMAGENET1K\_V2}) rather than training from random initialization. Despite the substantial domain gap between natural images and RHEED patterns, the pre-trained features provided a markedly better starting point: across the train/holdout splits, validation error fell far more rapidly during the first epochs and converged to a lower held-out error than scratch-initialized models (mean test MAE 3.1 vs 7.7). We therefore adopted ImageNet initialization throughout, as it improved both convergence speed and final performance at no additional cost. This is consistent with reports that pre-trained networks substantially reduce the amount of labeled data required in scientific-imaging tasks \cite{Stuckner_2022}, including RHEED-based growth analysis \cite{Price_2024}.

To ensure that our results are not biased by a particular train/holdout split, we report outcomes across all possible structure‑level partitions. Consequently, more results naturally arise from smaller values of $p$ (see {\bf Dataset Construction}).

Rather than the canonical SGD-with-momentum recipe used to train ResNet on ImageNet \cite{He_2016}, we adopted a standard AdamW setup \cite{loshchilov2019} with a learning rate of $10^{-3}$ (the AdamW default) and a weight decay of $10^{-4}$, held fixed across all experiments. No systematic hyperparameter search was performed. Our effort was instead directed at the architectural and data-representation choices described above (input resolution, ImageNet initialization, augmentation, the number of validation structures, and the train/holdout ratio), which is where the observed performance gains originate. The results should therefore be read as the outcome of a single, untuned training configuration, and a dedicated hyperparameter search would likely yield further improvements. 

Although the loss function was based on the MSE for each fold, the final model was selected as the checkpoint with the lowest validation MAE. MAE was chosen as the primary metric representing model performance because it directly corresponds to the intercept parameter - to be introduced in the next methodology step - which quantifies the inference error after postprocessing.
 
    \item  {\bf Physical constraint from acquisition}
    
    During data acquisition, the sample rotates at a constant angular velocity. Thus, if the first image 
    $P_0$ corresponds to an unknown angle $\alpha_0$, subsequent angles satisfy
    \begin{equation}
    \alpha_k=\alpha_0+k\Delta\alpha
    \end{equation}
    where $\Delta\alpha$ is derived from the rotation velocity.
    \item {\bf Ideal NN predictions}
    
    In the ideal case where $NN(P_k)-\alpha_k=0$ for all $k$ 
    the plot of $NN(P_k)$ versus $\alpha_k$ forms a straight line with slope +1 or 
-1, depending on the rotation direction (i.e. sign of $\Delta\alpha$). The only unknown parameter is $\alpha_0$ which can be extracted from the intercept of this line. The ideal relationship is therefore,
\begin{equation}
   NN(P_k)=k\Delta\alpha+\alpha_0 
\end{equation}

\item {\bf Realistic NN predictions}

In practice, the network’s predictions deviate from the ideal linear trend, producing a noticeably noisier pattern, as illustrated in Figure~\ref{P-NN-vs-target} by the points scattered around the ideal relationship. For the example shown, the MAE corresponds to an average deviation of approximately $\pm 1.77^\circ$, which is already below the error tolerance required for practical use. Individual predictions, however, may exhibit substantially larger errors, as demonstrated by the clearly visible outliers. What is more, the inference of the 2D neural network does not provide any information on the direction of sample rotation. 

\begin{figure}[h]
    \centering
    \includegraphics[width=0.45\textwidth]{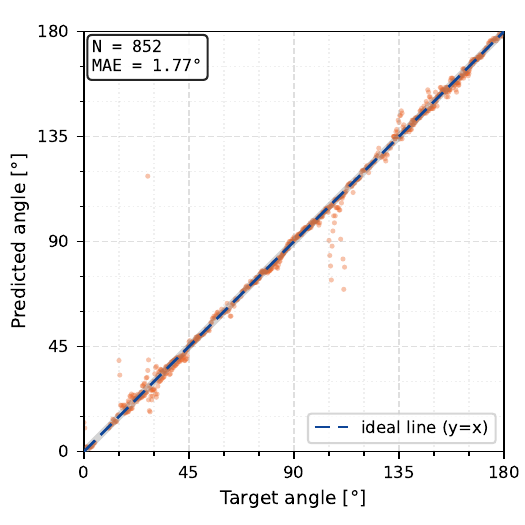}
    \caption{Visualization of $\alpha_k$ vs.\ NN($P_k$) in the realistic case for N=852 RHEED pictures of a single RHEED rotation (scattered points), compared with the ideal linear trend (blue line). }
    \label{P-NN-vs-target}
\end{figure}
Therefore there are two main challenges that the postprocessing needs to address: 

\noindent{\it(i)} determining the correct sign of the slope (i.e., the sign of $\Delta \alpha$), and

\noindent{\it(ii)} estimating the intercept $\alpha_0$ as accurately as possible.

An additional issue worth mentioning is that prediction errors may be biased, leading to an asymmetric distribution of residuals, $NN(P_k)-k\Delta\alpha$ (cf. Figure~\ref{P-NN-vs-target}), around the desired value $\alpha_0$.

\item {\bf Slope determination}

Because the neural network is trained only on images corresponding to angles in the range $0-180^\circ$, its predictions are effectively restricted to this interval. Consequently, if the true physical angle $\alpha_k=k\Delta\alpha+\alpha_0$ moves outside this range, the NN cannot continue the linear trend and instead produces values that “wrap around” to remain within $0-180^\circ$.
For example, if $\alpha_0=100^\circ$ and $k\Delta\alpha=80^\circ$ the true angle reaches $180^\circ$ and then NN will output values near $0^\circ$. This produces a characteristic drop (or rise) in the predicted sequence. 

To address this, predictions are processed in time windows of angular width $\Delta\alpha_{\rm win}$ (default $180^\circ$). 
 For the first window, two hypotheses are tested: $s = +1$ and $s = -1$. For each, modular residuals 
 \begin{equation}
  r_k^{(\pm)} = \left(NN(P_k) \mp (k\Delta\alpha \bmod 180)\right) \bmod 180   
 \end{equation}
 are computed, followed by their circular concentration 
 \begin{equation}
 R = \left|\operatorname{mean}\left(\exp\left(2\pi i \cdot r_k^{(\pm)}/180\right)\right)\right|   
 \end{equation}
 The hypothesis with the higher concentration corresponds to the correct orientation; the selected sign is stored and applied to all subsequent windows.

\item {\bf Robust estimation of the intercept}\\
Once the sign of the slope is fixed, we compute the modular residuals 
\begin{equation}
r_k = (NN(P_k) - s \cdot (k\Delta\alpha \bmod 180)) \bmod 180    
\end{equation}
for every frame in the window. Since the network was trained on the range $[0^\circ, 180^\circ)$, all subsequent processing is carried out in modular arithmetic with period $180^\circ$.

Before rejecting outliers, the residuals must be re-centered to remove the discontinuities introduced by modular wrap-around. We compute their circular mean (from the argument of the mean of the complex vectors $\exp(2\pi i \cdot r_k/180)$) and shift the residuals so that this mean sits at the center of the interval, at $90^\circ$. This prevents points that are close in the circular sense from being incorrectly treated as outliers.

We then estimate the intercept iteratively on the re-centered residuals. At each step we estimate the mode of the residual distribution using kernel density estimation (KDE) and discard points that satisfy $|r_k - \mathrm{mode}| > \delta$. Two iterations are used, with $\delta = 10^\circ$ and $\delta = 5^\circ$, respectively. After cleaning, the circular mean of the original (unshifted) retained residuals is taken as the intercept estimate $\hat{\alpha}_0^{(z)}$ for the $z$-th window.

The final estimate $\hat{\alpha}_0$ is the circular mean of the intercepts from all windows processed so far:
\[
\hspace{2em}
\hat{\alpha}_0 = \frac{180}{2\pi}
\arg\!\left( \frac{1}{K} \sum_{z=1}^{K}
\exp\!\left( \tfrac{2\pi i\, \hat{\alpha}_0^{(z)}}{180} \right) \right)
\bmod 180^\circ,
\]
where $K$ is the number of windows in which enough observations remained after outlier removal to estimate the crystallographic direction angle.

\end{enumerate}

\subsection{3D ResNet approach}
\label{sec:resnet3d}

Although the methodology presented in the previous subsection is central to this work, for benchmarking purposes we also report results obtained using a 3D ResNet architecture. When there is a wealth of training data and strong computational resources are available, ResNet3D is, in principle, well suited to the problem at hand.

In this approach, several consecutive RHEED images, separated by a fixed angular interval, are stacked into a three‑dimensional tensor and fed into a 3D ResNet neural network. This allows the model to infer the direction of rotation directly from the temporal structure present in the input sequence. A related use of deep learning on sequences of in-situ images has recently proven effective for real-time growth diagnostics in pulsed-laser deposition \cite{Harris_2024_plume}. Moreover, unlike the previous method - which requires a full rotation and the complete set of acquired images to perform effective inference - the 3D ResNet can already provide meaningful predictions from only the first few RHEED frames.

Formally for this approach we create a new data set with $\{\{P_{k},P_{k+\nu},P_{k+2\nu},...,P_{k+\chi\nu}\},\alpha_{k+m}\}$ where
$\nu$ constitutes the angle shift between neighboring RHEED pictures in the data set (can be positive and negative), $\chi+1$ is the number of pictures in the single data set, and $\alpha_{k+m}$ is the label of the frame chosen as the inference target, $m \in \{0, \nu, ..., \chi\nu\}$. Both $\nu$ and $\chi$ can be considered hyperparameters. Here however for benchmark purposes those parameters have been fixed to $\chi = 3$ and $\nu = 0.33^\circ$, which means that we take 4 neighbouring frames, and we set $m = \chi\nu$, i.e. the last frame in the sequence serves as the inference target. 

For the 3D ResNet architecture, we explored a range of hyperparameters to identify configurations that provide stable training and good performance. Using the AdamW optimizer, we compared different values of the weight-decay coefficient and several learning-rate scheduling strategies, focusing on Cosine Annealing with Warm Restarts while varying both the restart period and its multiplicative expansion factor. We further examined the input resolution, the number of consecutive RHEED frames used as input (clip length) and their temporal stride, and whether augmentation was applied. Throughout, the network was trained by minimizing the MSE, and for each fold, the final model was selected as the checkpoint with the lowest validation MAE.

The goal of trying ResNet3D is to train it to reach a performance comparable to that of the best 2D ResNet model equipped with the postprocessing pipeline. By directly contrasting the two approaches, we shall demonstrate that the 2D model with postprocessing achieves similar predictive performance while requiring substantially less training data and significantly lower computational resources. This comparison highlights the efficiency gains introduced by the postprocessing step and shows that a well‑designed 2D pipeline can match the performance of a considerably more expensive 3D architecture.

  \begin{figure*}[htbp]
    \centering
    \includegraphics[width=0.45\textwidth]{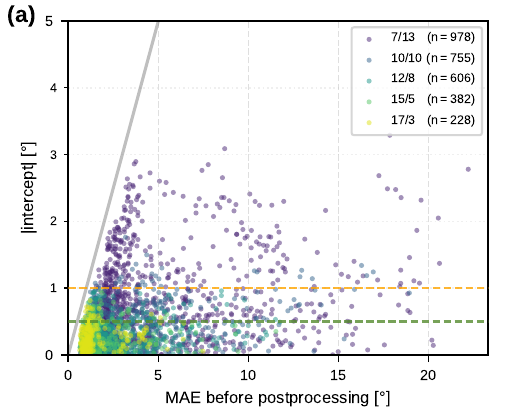}
    \includegraphics[width=0.45\textwidth]{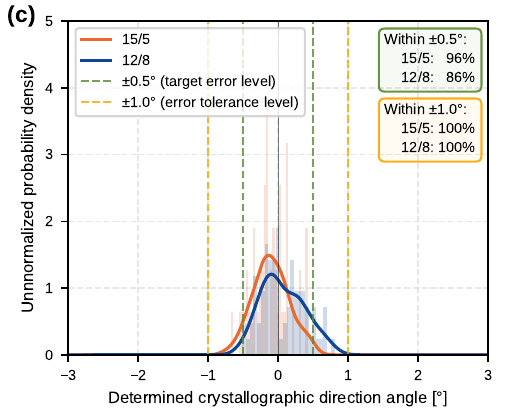}
    \includegraphics[width=0.45\textwidth]{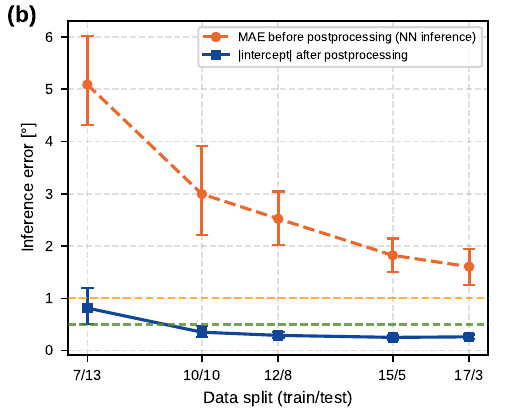}
    \includegraphics[width=0.45\textwidth]{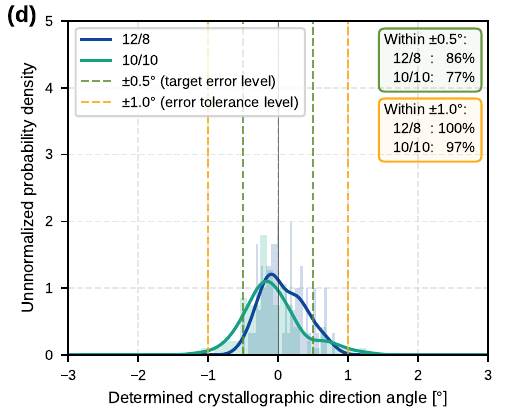}
   \caption{
\textbf{(a)} Distribution of test--set rotations plotted in a two--dimensional space, where each point corresponds to a single rotation and is positioned according to its MAE before postprocessing (x--axis) and the intercept obtained after postprocessing (y--axis). The grey line denotes the $x = y$ reference; points below this line indicate improvement due to postprocessing.
\textbf{(b)} Mean target--estimation error - MAE before postprocessing and absolute value of intercept after postprocessing - across different data splits. Bars denote 95\% bootstrap confidence intervals of the mean. Lines connecting points serve only as a guide for the eye.
\textbf{(c,d)} Unnormalized probability density of the intercept estimates for all single--rotation test samples, shown for the postprocessed neural--network inference across selected data splits.
\textbf{(a--d)} In all panels, the yellow and green dashed lines indicate error tolerance ($\pm 1.0^\circ$) and target error ($\pm 0.5^\circ$) levels for the intercept obtained after postprocessing. 
}
    \label{intercept_stats}
\end{figure*}

\section{Results and discussion}
\label{sec:results}
\subsection{ResNet2D}
\label{sec:results2d}

\begin{table}[ht]
  \centering
  \renewcommand{\arraystretch}{1.4}
  \caption{Training configuration for the 2D ResNet-50 angle regressor;
           the same settings were used for every fold.}
  \resizebox{\columnwidth}{!}{%
  \begin{tabular}{l l}
    \hline
    Setting & Value \\
    \hline
    Backbone        & \makecell[l]{ResNet-50, ImageNet \texttt{IMAGENET1K\_V2}\\(all layers trained)} \\
    Regression head & single linear output unit (scalar) \\
    Input           & \makecell[l]{$56\times96$ grayscale, replicated\\to 3 channels, standardized} \\
    Optimizer       & AdamW \\
    Learning rate   & $10^{-3}$ \\
    Weight decay    & $10^{-4}$ \\
    LR schedule     & \makecell[l]{cosine annealing, 5-epoch linear\\warmup, $\eta_{\min}{=}10^{-6}$} \\
    Epochs          & 150 \\
    Batch size      & 2048 (train) / 512 (validation) \\
    Loss            & mean squared error \\
    Precision       & bfloat16 mixed precision \\
    Augmentation    & \makecell[l]{on-the-fly, re-sampled\\per access (cf.\ Figure~\ref{on-the-fly-augmentation})} \\
    Model selection & lowest validation-MAE checkpoint \\
    \hline
  \end{tabular}}
  \label{tab:training}
\end{table}

The full training configuration is summarized in Table~\ref{tab:training}. We trained the network to minimize the MSE with the AdamW optimizer under a cosine learning-rate schedule with a short linear warmup, using the on-the-fly augmentation introduced above (cf. Figure~\ref{on-the-fly-augmentation}), re-sampled at every access. Every fold was trained for 150 epochs, and we retained the checkpoint with the lowest validation MAE for evaluation. 

First, we evaluated two RHEED image resolutions and found that for 2D ResNet without postprocessing they yield equivalent holdout performance: mean MAE of $1.48^\circ$ for $96 \times 56$ versus $1.44^\circ$ for $112 \times 64$. The $0.04^\circ$ difference between resolutions is negligible compared to the per-fold cross-validation variability ($\pm 0.5^\circ$). Consequently, we selected the lower resolution ($96 \times 56$) for training the 2D ResNet to improve computational efficiency.

Further, to mitigate checkpoint sensitivity to fluctuations in the validation loss, we evaluated increasing the validation set size from 2 to 3 structures. While a larger validation set effectively prevents a single atypical structure from skewing epoch selection, reducing the training set size by $1/(p-2)$ led to a noticeable performance degradation. Specifically, in our setting, this reduction resulted in an increase in mean holdout MAE of approximately $0.2^\circ$ across all folds. Consequently, to maximize training capacity while preserving generalization performance, the validation size was fixed at 2 structures for all reported splits.

In Figure~\ref{intercept_stats} we holistically present the results for different splits between the training and holdout datasets, collectively illustrating how well the 2D architecture is able to infer the rotation angle even when trained on a limited amount of data. For visualization, the target variable - the initial sample orientation - was always set to $0^\circ$. The dashed green and yellow lines, indicating the target error level ($\pm 0.5^\circ$) and error tolerance level ($\pm 1^\circ$), respectively, appear consistently across all four subpanels of Figure~\ref{intercept_stats}. The target error level reflects the accuracy with which an experienced grower can reliably judge the growth on a given crystallographic direction, and is close to the experimental frame-collection resolution ($0.33^\circ$). The error tolerance level marks the boundary at which the inferred direction is still practically useful: a small manual correction toward the proper direction may be needed, yet growth-related decisions can still be made.

In Figure~\ref{intercept_stats}a) we compare the results of the bare 2D ResNet - summarized by MAE for each run and each train/holdout split ($p/20-p$) - with the results obtained after postprocessing, where the inferred crystallographic direction is estimated over the full parameter space (with target $0^\circ$). For visual guidance, we include a grey line representing the relation $\text{intercept} = \text{MAE}$ (note the different scales on the horizontal and vertical axes). Points universally lying below this line indicate that postprocessing improves the inference quality. Although the largest improvement occurs for the 7/13 split (a reduction from $\text{MAE} \simeq 22^\circ$ to $\text{intercept} \simeq 3^\circ$), the target error and error tolerance levels seem to be consistently reached only for the 15/5 and 17/3 splits.

To highlight the boundary clearly, Figure~\ref{intercept_stats}b) shows the average performance across all runs for each split, both for the bare 2D network (measured by MAE) and for the postprocessed network (measured by the intercept). Because multiple samplings correspond nominally to the same split $p/20-p$, we also plot bars representing 95\% bootstrap confidence intervals to demonstrate the distribution of the error. It is evident that the bare network does not reach the error tolerance level for any split, whereas the postprocessed network remains consistently below the target error level already for the 10/10 split.

To further characterize the distribution of scores produced by the clearly superior postprocessed network, Figure~\ref{intercept_stats}c–d show pairwise comparisons of the corresponding probability densities for the 15/5 vs.\ 12/8 splits and for the 12/8 vs.\ 10/10 splits. From these comparisons we conclude that, for the 15/5 split (i.e., using only 15 structures for training), the postprocessed neural network predicts the target variable within the error tolerance level for essentially all cases, and in $96\%$ of cases remains within the target error level - close to the experimental frame-collection resolution. This allows us to conclude that the neural-powered solution presented here constitutes an effective and universally accurate tool for predicting crystallographic directions from RHEED images.

The trained ResNet2D models were additionally evaluated on CPU on the same machine used for RHEED data acquisition (Table \ref{tbl:hardware}). Inference ran comfortably at 10 FPS, enabling real‑time estimation of the crystallographic direction from incoming RHEED frames. Beyond playback tests, the model was also run on the live RHEED feed of the system during growth; using its output to automatically stop the substrate rotation is planned for forthcoming experiments. This frame rate is sufficient for using the neural network as a control mechanism to stop the MBE substrate rotation during actual growth, serving as a preparatory step for subsequent experimental procedures. Example inference runs on recorded RHEED videos - including structures grown outside the training-data period - are provided as Supplementary Videos and described in the Supporting Information\cite{supp_info}.
 
\subsection{ResNet3D}
\label{sec:results3d}

For a clear comparison, the training of the 3D ResNet models is done on the same resolution of pictures as prepared for the 2D ResNet model. Training was performed using the MSE loss function and the AdamW optimizer with a weight-decay coefficient of $1\times 10^{-5}$. The learning rate followed a Cosine Annealing with Warm Restarts schedule, with an initial period of $T=50$, a period-doubling factor of 2, an initial learning rate of $2\times 10^{-3}$, and a minimum learning rate of $1\times 10^{-7}$. The model was trained on sequences of four consecutive RHEED frames (0.33$^\circ$ spacing), using the same augmentation pipeline as in the 2D experiments, and the inference target was the last frame in each sequence. Training ran for 150 epochs and required approximately 11 hours on a single Nvidia A6000 GPU (48\,GB), with 8 data-loading workers and prefetching enabled. Final evaluation was conducted using MAE.

\begin{figure}[h]
    \centering
    \includegraphics[width=0.45\textwidth]{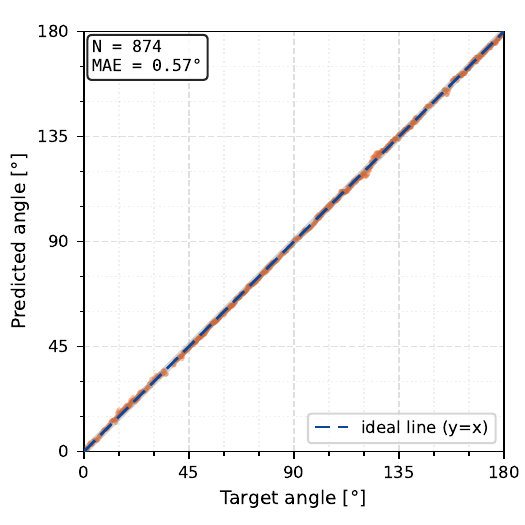}
    \caption{Visualization of $\alpha_{k+3}$ vs.\ $NN_{3D}$($\{P_k,...,P_{k+3}\}$) in the representative case for N=874 RHEED pictures of a single RHEED rotation (scattered points), compared with the ideal linear trend (blue dashed line). }
    \label{P-NN-vs-target_3D}
\end{figure}

As we described in Section \ref{sec:methodology}, we test only one train/holdout data split for ResNet3D, i.e. 19/1, and a single 3D picture needed for the inference is composed of four consecutive frames from a single rotation. Again we follow the schema in Figure~\ref{dataset_division}, i.e. we sample the holdout structure 10 times and perform training on the remaining 19 structures. In Figure~\ref{P-NN-vs-target_3D} we show the inference of the bare ResNet3D. When compared to the analogous Figure~\ref{P-NN-vs-target} for ResNet2D, it is clear that ResNet3D no longer suffers from outliers, and the MAE is already a reliable measure of inference quality at any angle.

\begin{figure}[h]
    \centering
    \includegraphics[width=0.45\textwidth]{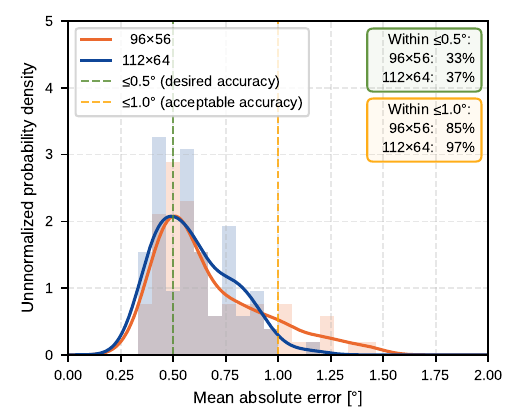}
    \caption{Unnormalized probability density of the MAE for all
single–rotation test samples, shown for the ResNet3D neural–network inference across two pictures resolution used in training. The yellow and green vertical dashed lines indicate error tolerance ($\pm1.0^\circ$) and target error ($\pm0.5^\circ$) levels of main crystallographic direction estimation.}
    \label{MAE_intercept_3D}
\end{figure}

In Figure~\ref{MAE_intercept_3D} we show the distribution of the unnormalized probability for two chosen initial picture resolutions used during training. For the resolution 96×56, the probability mass is centered within the error tolerance region (though not within the target error region), and 15\% of cases fall outside the error tolerance boundary. The performance can be slightly improved by increasing the resolution to 112×64, at the cost of approximately 25\%
 longer training time. With this higher resolution, the mean MAE (computed over all rotations from the holdout structure) shifts slightly closer to the target 0$^\circ$ angle, although it still does not reach the target error level. On the other hand, the error dispersion decreases, and now only $3\%$ of cases fall beyond the target error region. For comparison, see the first column of Table~\ref{tab1}.
 
Naturally, there is no obstacle to enriching the inference of ResNet3D with the postprocessing defined for ResNet2D. The performance, measured by the value of the intercept, is presented in the second column of Table~\ref{tab1}. It is evident that the performance is significantly improved: the mean across all rotations from the test structure remains within the target error boundary, and the error dispersion is minimal. Moreover, there is no significant difference between the two picture resolutions used during training.

 \begin{table}[ht]
  
 \renewcommand{\arraystretch}{1.1}
 \caption{Comparison of the bare ResNet3D model (evaluated using MAE) and the ResNet3D model augmented with the postprocessing pipeline (evaluated using the intercept), across two input image resolutions. The target angle is $0^\circ$. Training time includes only the model training; postprocessing is measured in seconds and its contribution to the total time is negligible. Note that errors in the table are provided with the standard deviation.}
  \begin{tabular}{cccc}
    \hline
    Resolution & MAE  & Intercept & time \\
    \hline
    $96 \times 56$  &   0.771 $\pm$ 0.532     &   0.205 $\pm$ 0.071 &11 h 2 m\\
    $112 \times 64$  &  0.684 $\pm$ 0.373     &   0.204 $\pm$ 0.048& 14 h 9m\\
    \hline
  \end{tabular}
  
  \label{tab1}
\end{table}

\subsection{Comparison between ResNet2D and ResNet3D}
\label{sec:comparison}

In this subsection we compare the performance of our pipeline when powered by ResNet2D and ResNet3D. Table~\ref{tab:reg_metrics} summarizes the key characteristics of both approaches, including the computational cost (measured by training time), the amount of training data required (number of structures), and the inference quality (measured either by MAE for the bare models or by the intercept when postprocessing is applied).

The main conclusion is that, thanks to the dedicated postprocessing pipeline, the moderately trained ResNet2D achieves performance superior to the bare ResNet3D and comparable to the postprocessed ResNet3D, while requiring substantially less training data. This is particularly important in the context of RHEED images from MBE growth, where data availability is a critical bottleneck. Moreover, the training time for ResNet2D is significantly shorter, making it a more practical choice when computational resources or data volume are limited.

We note that the two approaches differ in inference latency. The postprocessing pipeline requires predictions accumulated over a full angular window of $\Delta\alpha_{\rm win} = 180^\circ$ - about 30 s of acquisition at 1 RPM - before the first reliable estimate of the intercept becomes available, whereas the bare ResNet3D provides meaningful predictions already from the first $\chi+1$ frames. The comparison in Table~\ref{tab:reg_metrics} should be read with this distinction in mind: the postprocessed ResNet2D is preferable whenever a fraction of a rotation of delay is acceptable, while ResNet3D minimizes the time to the first estimate at the cost of training time and data requirements.

\begin{table}[ht]
  \centering
  \caption{Comparison of ResNet2D and ResNet3D across training time, amount of training data, and inference error for input resolution 96×56. MAE is reported for the bare ResNet3D, while intercept reflects performance after applying the postprocessing pipeline. The superscript * next to ResNet2D and ResNet3D denotes that postprocessing is used. Note that errors in the table are provided with the standard deviation.}
  \resizebox{\columnwidth}{!}{
\begin{tabular}{c|ccc}
\hline
 & ResNet2D* & ResNet3D & ResNet3D* \\
\hline
Train time  & 33 m 57 s & 11 h 2 m & 11 h 2 m \\
\# train structures & 15 & 19 & 19 \\
\makecell{
{Estimate error (of target}\\ 
{$0^\circ$) on test set}} 
 & \makecell{0.217 $\pm$ 0.157 \\ (intercept)} 
 & \makecell{0.771 $\pm$ 0.532 \\ (MAE)}
 & \makecell{0.205 $\pm$ 0.071 \\ (intercept)} \\
\hline
\end{tabular}}

  \label{tab:reg_metrics}
\end{table}
\section{Summary}
\label{sec:summary}
In this work we developed a data‑efficient neural‑vision pipeline for automated crystallographic alignment during MBE growth, addressing the practical challenge of limited availability of diverse RHEED datasets. By systematically evaluating both 2D and 3D ResNet architectures, each with and without physics‑aware postprocessing, we demonstrated that reliable inference of crystallographic directions can be achieved even when training data and computational resources are limited.

As a benchmarking reference, we considered the bare ResNet3D model, which benefits from the temporal information contained in short RHEED sequences. It produces stable predictions without outliers and yields MAE values that remain meaningful across the full rotation range. However, this stability comes at the cost of long training times and a gap between the raw MAE and the target error boundary with the available data. Increasing the input resolution improves performance only marginally and does not eliminate the need for further refinement.

In contrast, and constituting the main result of this work, we show that the ResNet2D architecture, when combined with the dedicated postprocessing pipeline, achieves performance superior to the bare ResNet3D and comparable to the postprocessed ResNet3D, while requiring substantially fewer training structures and more than an order of magnitude shorter training time. The postprocessing step proves essential: it consistently reduces prediction error across all train/holdout splits, enabling the 2D model to reach the desired performance even when trained on as few as 10–15 structures. Probability‑density analyses confirm that for the 15/5 split, the postprocessed network predicts the crystallographic direction within the error tolerance region for essentially all cases and within the target error boundary for 96\% of cases - approaching the experimental resolution of RHEED frame acquisition.

Beyond performance, the 2D model also satisfies practical constraints. When deployed on the machine used for RHEED acquisition, inference runs at 10 FPS on a CPU‑only setup, enabling real‑time estimation of crystallographic orientation and providing a viable control mechanism for stopping substrate rotation during MBE growth.

Overall, this study demonstrates that carefully designed physics‑aware postprocessing can compensate for limited training data and allow lightweight architectures to match the performance of more complex models. The resulting system is fully trained and ready for deployment in future CdTe growth experiments, and the methodology is designed to transfer to other materials systems where RHEED data are limited. More broadly, our work highlights the potential of machine‑learning‑assisted automation in thin‑film synthesis and represents a step toward AI‑driven MBE growth.

\section*{Acknowledgments}
This research was supported by the “MagTop” project (FENG.02.01-IP.05-0028/23) carried out within the “International Research Agendas” programme of the Foundation for Polish Science co-financed by the European Union under the European Funds for Smart Economy 2021-2027 (FENG).

Publication subsidized from the state budget within the framework of the programme of the Minister of Science (Polska) called Polish Metrology II project no. PM-II/SP/0012/2024/02, amount of subsidy 944,900.00 PLN, total value of the project 944,900.00 PLN.

\section*{Data availability}
The RHEED dataset used in this work is openly available in the Zenodo repository at \url{https://doi.org/10.5281/zenodo.21534842}.

\bibliographystyle{apsrev4-2}
\bibliography{bibliography}

@article{He_2016,
    title={Deep Residual Learning for Image Recognition},
    author={Kaiming He and Xiangyu Zhang and Shaoqing Ren and Jian Sun},
    journal={Proceedings of the IEEE Conference on Computer Vision and Pattern Recognition (CVPR)},
    pages={770--778},
    year={2016},
    doi={10.1109/CVPR.2016.90},
    url={https://doi.org/10.1109/CVPR.2016.90}
}

@article{Xie_2017,
    title={Aggregated Residual Transformations for Deep Neural Networks},
    author={Saining Xie and Ross Girshick and Piotr Doll{\'a}r and Zhuowen Tu and Kaiming He},
    journal={Proceedings of the IEEE Conference on Computer Vision and Pattern Recognition (CVPR)},
    pages={5987--5995},
    year={2017},
    doi={10.1109/CVPR.2017.634},
    url={https://doi.org/10.1109/CVPR.2017.634}
}

@article{Liu_2022_ConvNeXt,
    title={A ConvNet for the 2020s},
    author={Zhuang Liu and Hanzi Mao and Chao-Yuan Wu and Christoph Feichtenhofer and Trevor Darrell and Saining Xie},
    journal={Proceedings of the IEEE/CVF Conference on Computer Vision and Pattern Recognition (CVPR)},
    pages={11976--11986},
    year={2022},
    doi={10.1109/CVPR52688.2022.01167},
    url={https://doi.org/10.1109/CVPR52688.2022.01167}
}

@article{Tan_2021_EfficientNetV2,
    title={EfficientNetV2: Smaller Models and Faster Training},
    author={Mingxing Tan and Quoc V. Le},
    journal={Proceedings of the 38th International Conference on Machine Learning (ICML)},
    pages={10096--10106},
    year={2021},
    url={https://proceedings.mlr.press/v139/tan21a.html}
}

@article{Dosovitskiy_2021,
    title={An Image is Worth 16×16 Words: Transformers for Image Recognition at Scale},
    author={Alexey Dosovitskiy and Lucas Beyer and Alexander Kolesnikov and Dirk Weissenborn and Xiaohua Zhai and Thomas Unterthiner and Mostafa Dehghani and Matthias Minderer and Georg Heigold and Sylvain Gelly and Jakob Uszkoreit and Neil Houlsby},
    journal={International Conference on Learning Representations (ICLR)},
    year={2021},
    url={https://openreview.net/forum?id=YicbFdNTTy}
}

@article{Liu_2021_Swin,
    title={Swin Transformer: Hierarchical Vision Transformer Using Shifted Windows},
    author={Ze Liu and Yutong Lin and Yue Cao and Han Hu and Yixuan Wei and Zheng Zhang and Stephen Lin and Baining Guo},
    journal={Proceedings of the IEEE/CVF International Conference on Computer Vision (ICCV)},
    pages={10012--10022},
    year={2021},
    doi={10.1109/ICCV48922.2021.00986},
    url={https://doi.org/10.1109/ICCV48922.2021.00986}
}

@article{Touvron_2021_DeiT,
    title={Training Data-Efficient Image Transformers and Distillation through Attention},
    author={Hugo Touvron and Matthieu Cord and Alexandre Sablayrolles and Gabriel Synnaeve and Herv{\'e} J{\'e}gou},
    journal={Proceedings of the 38th International Conference on Machine Learning (ICML)},
    pages={10347--10357},
    year={2021},
    url={https://proceedings.mlr.press/v139/touvron21a.html}
}

@article{Shen2024,
    author = {Shen, Chao and Zhan, Wenkang and Tang, Jian and Wu, Zhaofeng and Xu, Bo and Zhao, Chao and Wang, Zhanguo},
    title = {Universal Deoxidation of Semiconductor Substrates Assisted by Machine Learning and Real-Time Feedback Control},
    journal = {ACS Applied Materials \& Interfaces},
    volume = {16},
    number = {14},
    pages = {18213-18221},
    year = {2024},
    doi = {10.1021/acsami.4c01765},
    note ={PMID: 38554077},
    URL = {https://doi.org/10.1021/acsami.4c01765},
}

@article{Yu2025,
	author = {Yu, Mingyu and Moses, Isaiah A. and Reinhart, Wesley F. and Law, Stephanie},
	title = {Multimodal Machine Learning Analysis of GaSe Molecular Beam Epitaxy Growth Conditions},
	journal = {ACS Applied Materials \& Interfaces},
	volume = {17},
	number = {23},
	pages = {34707-34716},
	year = {2025},
	doi = {10.1021/acsami.5c02891},
	note ={PMID: 40434265},
	URL = {https://doi.org/10.1021/acsami.5c02891},
}

@article{Chin_2025,
    title = {Analyzing the impact of Se concentration during the molecular beam epitaxy deposition of 2D SnSe with atomistic-scale simulations and explainable machine learning},
    journal = {Materials Today Advances},
    volume = {28},
    pages = {100640},
    year = {2025},
    issn = {2590-0498},
    doi = {https://doi.org/10.1016/j.mtadv.2025.100640},
    url = {https://www.sciencedirect.com/science/article/pii/S2590049825000852},
    author = {Jonathan R. Chin and Isaiah A. Moses and Mengyi Wang and Marshall B. Frye and Mingyu Yu and Nadire Nayir and Maria Hilse and Adri C.T. {van Duin} and Stephanie Law and Wesley Reinhart and Lauren M. Garten}
}

@article{Ohkubo_2021,
    title = {Realization of closed-loop optimization of epitaxial titanium nitride thin-film growth via machine learning},
    journal = {Materials Today Physics},
    volume = {16},
    pages = {100296},
    year = {2021},
    issn = {2542-5293},
    doi = {https://doi.org/10.1016/j.mtphys.2020.100296},
    url = {https://www.sciencedirect.com/science/article/pii/S2542529320301206},
    author = {I. Ohkubo and Z. Hou and J.N. Lee and T. Aizawa and M. Lippmaa and T. Chikyow and K. Tsuda and T. Mori}
}

@article{Kwoen_2022,
    title = {Multiclass classification of reflection high-energy electron diffraction patterns using deep learning},
    journal = {Journal of Crystal Growth},
    volume = {593},
    pages = {126780},
    year = {2022},
    issn = {0022-0248},
    doi = {https://doi.org/10.1016/j.jcrysgro.2022.126780},
    url = {https://www.sciencedirect.com/science/article/pii/S0022024822002688},
    author = {Jinkwan Kwoen and Yasuhiko Arakawa}
}

@article{Chong_2024,
    title = {Machine-learning-empowered identification of initial growth modes for 2D transition metal dichalcogenide thin films},
    journal = {Applied Surface Science},
    volume = {669},
    pages = {160547},
    year = {2024},
    issn = {0169-4332},
    doi = {https://doi.org/10.1016/j.apsusc.2024.160547},
    url = {https://www.sciencedirect.com/science/article/pii/S0169433224012601},
    author = {Minsu Chong and Tae Gyu Rhee and Yeong Gwang Khim and Min-Hyoung Jung and Young-Min Kim and Hu Young Jeong and Heung-Sik Kim and Young Jun Chang and Hyuk Jin Kim}
}

@article{Xiong_2024,
    title = {Indexing high-noise electron backscatter diffraction patterns using convolutional neural network and transfer learning},
    journal = {Computational Materials Science},
    volume = {233},
    pages = {112718},
    year = {2024},
    issn = {0927-0256},
    doi = {https://doi.org/10.1016/j.commatsci.2023.112718},
    url = {https://www.sciencedirect.com/science/article/pii/S0927025623007127},
    author = {Guoqing Xiong and Changxin Wang and Yu Yan and Lei Zhang and Yanjing Su}
    }

@article{ChaoShen_2024,
    doi = {10.1088/1674-4926/45/3/031301},
    url = {https://doi.org/10.1088/1674-4926/45/3/031301},
    year = {2024},
    month = {mar},
    publisher = {Chinese Institute of Electronics},
    volume = {45},
    number = {3},
    pages = {031301},
    author = {Shen, Chao and Zhan, Wenkang and Li, Manyang and Sun, Zhenyu and Tang, Jian and Wu, Zhaofeng and Xu, Chi and Xu, Bo and Zhao, Chao and Wang, Zhanguo},
    title = {Development of in situ characterization techniques in molecular beam epitaxy},
    journal = {Journal of Semiconductors}
}

@article{Gemperline_2025,
    author = {Gemperline, Patrick T. and Paudel, Rajendra and Vasudevan, Rama K. and Comes, Ryan B.},
    title = {Improvement of data analytics techniques in reflection high-energy electron diffraction to enable machine learning},
    journal = {Journal of Vacuum Science \& Technology A},
    volume = {43},
    number = {3},
    pages = {032701},
    year = {2025},
    month = {03},
    issn = {0734-2101},
    doi = {10.1116/6.0004400},
    url = {https://doi.org/10.1116/6.0004400},
}

@article{Provence_2020,
    title = {Machine learning analysis of perovskite oxides grown by molecular beam epitaxy},
    author = {Provence, Sydney R. and Thapa, Suresh and Paudel, Rajendra and Truttmann, Tristan K. and Prakash, Abhinav and Jalan, Bharat and Comes, Ryan B.},
    journal = {Physical Review Materials},
    volume = {4},
    issue = {8},
    pages = {083807},
    numpages = {11},
    year = {2020},
    month = {Aug},
    publisher = {American Physical Society},
    doi = {10.1103/PhysRevMaterials.4.083807},
    url = {https://link.aps.org/doi/10.1103/PhysRevMaterials.4.083807}
}

@article{Kwoen_2020,
    author = {Kwoen, Jinkwan and Arakawa, Yasuhiko},
    title = {Classification of Reflection High-Energy Electron Diffraction Pattern Using Machine Learning},
    journal = {Crystal Growth \& Design},
    volume = {20},
    number = {8},
    pages = {5289-5293},
    year = {2020},
    doi = {10.1021/acs.cgd.0c00506},
    URL = {https://doi.org/10.1021/acs.cgd.0c00506},
}

@article{walieh_2023,
    author = {Khaireh-Walieh, Abdourahman and Arnoult, Alexandre and Plissard, S{\'e}bastien and Wiecha, Peter R.},
    title = {Monitoring MBE Substrate Deoxidation via RHEED Image-Sequence Analysis by Deep Learning},
    journal = {Crystal Growth \& Design},
    volume = {23},
    number = {2},
    pages = {892-898},
    year = {2023},
    doi = {10.1021/acs.cgd.2c01132},
    URL = {https://doi.org/10.1021/acs.cgd.2c01132},
}

@article{kim_2023,
    author = {Kim, Hyuk Jin and Chong, Minsu and Rhee, Tae Gyu and Khim, Yeong Gwang and Jung, Min-Hyoung and Kim, Young-Min and Jeong, Hu Young and Choi, Byoung Ki and Chang, Young Jun},
    title = {Machine-learning-assisted analysis of transition metal dichalcogenide thin-film growth},
    journal = {Nano Convergence},
    volume = {10},
    pages = {10},
    year = {2023},
    doi = {https://doi.org/10.1186/s40580-023-00359-5}
}

@article{Wakabayashi_2019,
    author = {Wakabayashi, Yuki K. and Otsuka, Takuma and Krockenberger, Yoshiharu and Sawada, Hiroshi and Taniyasu, Yoshitaka and Yamamoto, Hideki},
    title = {Machine-learning-assisted thin-film growth: Bayesian optimization in molecular beam epitaxy of SrRuO3 thin films},
    journal = {APL Materials},
    volume = {7},
    number = {10},
    pages = {101114},
    year = {2019},
    month = {10},
    doi = {10.1063/1.5123019},
    url = {https://doi.org/10.1063/1.5123019},
}

@article{Anjum_2023,
    author = {Anjum, Sharjeel and Lee, Hyeong-Yeon and Noh, Hong-Kyun},
    title = {Rotation Error Detection of Gallium Nitride (GaN) Substrate in MBE Utilizing Ensemble Learning},
    journal = {Crystal Growth \& Design},
    volume = {23},
    number = {6},
    pages = {4138-4146},
    year = {2023},
    doi = {10.1021/acs.cgd.2c01544},
    URL = {https://doi.org/10.1021/acs.cgd.2c01544},
}

@article{Liang_2022,
    title = {Application of machine learning to reflection high-energy electron diffraction images for automated structural phase mapping},
    author = {Liang, Haotong and Stanev, Valentin and Kusne, Aaron Gilad and Tsukahara, Yuto and Ito, Kaito and Takahashi, Ryota and Lippmaa, Mikk and Takeuchi, Ichiro},
    journal = {Physical Review Materials},
    volume = {6},
    issue = {6},
    pages = {063805},
    numpages = {9},
    year = {2022},
    month = {Jun},
    publisher = {American Physical Society},
    doi = {10.1103/PhysRevMaterials.6.063805},
    url = {https://link.aps.org/doi/10.1103/PhysRevMaterials.6.063805}
}

@article{Shen_2024,
    title = {Machine-learning-assisted and real-time-feedback-controlled growth of InAs/GaAs quantum dots},
    author = {Shen, Chao and Zhan, Wenkang and Xin, Kaiyao and Li, Manyang and Sun, Zhenyu and Cong, Hui and Xu, Chi and Tang, Jian and Wu, Zhaofeng and Xu, Bo and Wei, Zhongming and Xue, Chunlai and Zhao, Chao and Wang, Zhanguo},
    journal = {Nature Communications},
    volume = {15},
    pages = {2724},
    year = {2024},
    doi = {10.1038/s41467-024-47087-w},
    url = {https://doi.org/10.1038/s41467-024-47087-w}
}

@article{Kaspar_2025,
    author = {Kaspar, Tiffany C. and Akers, Sarah and Sprueill, Henry W. and Ter-Petrosyan, Arman H. and Bilbrey, Jenna A. and Hopkins, Derek and Harilal, Ajay and Christudasjustus, Jijo and Gemperline, Patrick and Comes, Ryan B.},
    title = {Machine-learning-enabled on-the-fly analysis of RHEED patterns during thin film deposition by molecular beam epitaxy},
    journal = {Journal of Vacuum Science \& Technology A},
    volume = {43},
    number = {3},
    pages = {032702},
    year = {2025},
    month = {03},
    issn = {0734-2101},
    doi = {10.1116/6.0004493},
    url = {https://doi.org/10.1116/6.0004493},
}

@article{Keys1981,
  author={Keys, R.},
  journal={IEEE Transactions on Acoustics, Speech, and Signal Processing}, 
  title={Cubic convolution interpolation for digital image processing}, 
  year={1981},
  volume={29},
  number={6},
  pages={1153-1160},
  doi={10.1109/TASSP.1981.1163711}
}

@misc{loshchilov2019,
      title={Decoupled Weight Decay Regularization}, 
      author={Ilya Loshchilov and Frank Hutter},
      year={2019},
      eprint={1711.05101},
      archivePrefix={arXiv},
      primaryClass={cs.LG},
      doi={10.48550/arXiv.1711.05101}, 
}

@article{Polaczynski_2024,
    title = {3D topological semimetal phases of strained α-Sn on insulating substrate},
    journal = {Materials Today},
    volume = {75},
    pages = {135-148},
    year = {2024},
    issn = {1369-7021},
    doi = {https://doi.org/10.1016/j.mattod.2024.04.014},
    url = {https://www.sciencedirect.com/science/article/pii/S1369702124000816},
    author = {Jakub Polaczyński and Gauthier Krizman and Alexandr Kazakov and Bartłomiej Turowski and Joaquín Bermejo Ortiz and Rafał Rudniewski and Tomasz Wojciechowski and Piotr Dłużewski and Marta Aleszkiewicz and Wojciech Zaleszczyk and Bogusława Kurowska and Zahir Muhammad and Marcin Rosmus and Natalia Olszowska and Louis-Anne {de Vaulchier} and Yves Guldner and Tomasz Wojtowicz and Valentine V. Volobuev}
}

@Article{Carmody2012,
    author={Carmody, M. and Yulius, A. and Edwall, D. and Lee, D. and Piquette, E. and Jacobs, R. and Benson, D. and Stoltz, A. and Markunas, J. and Almeida, A. and Arias, J.},
    title={Recent Progress in MBE Growth of CdTe and HgCdTe on (211)B GaAs Substrates},
    journal={Journal of Electronic Materials},
    year={2012},
    month={Oct},
    day={01},
    volume={41},
    number={10},
    pages={2719-2724},
    doi={10.1007/s11664-012-2129-z},
    url={https://doi.org/10.1007/s11664-012-2129-z}
    }

@Article{Pan2022,
    author={Pan, W. W. and Gu, R. J. and Zhang, Z. K. and Lei, W. and Umana-Membreno, G. A. and Smith, D. J. and Antoszewski, J. and Faraone, L.},
    title={Defect Engineering in MBE-Grown CdTe Buffer Layers on GaAs (211)B Substrates},
    journal={Journal of Electronic Materials},
    year={2022},
    month={Sep},
    day={01},
    volume={51},
    number={9},
    pages={4869-4883},
    doi={10.1007/s11664-022-09725-1},
    url={https://doi.org/10.1007/s11664-022-09725-1}
    }

@article{VVV2017,
    author = {Volobuev, Valentine V. and Mandal, Partha S. and Galicka, Marta and Caha, Ondřej and Sánchez-Barriga, Jaime and Di Sante, Domenico and Varykhalov, Andrei and Khiar, Amir and Picozzi, Silvia and Bauer, Günther and Kacman, Perla and Buczko, Ryszard and Rader, Oliver and Springholz, Gunther},
    title = {Giant Rashba Splitting in Pb1–xSnxTe (111) Topological Crystalline Insulator Films Controlled by Bi Doping in the Bulk},
    journal = {Advanced Materials},
    volume = {29},
    number = {3},
    pages = {1604185},
    doi = {https://doi.org/10.1002/adma.201604185},
    url = {https://doi.org/10.1002/adma.201604185},
    year = {2017}
}

@article{kazakov2025,
  title = {Topological phase diagram and quantum magnetotransport effects in (Pb,Sn)Se quantum wells with magnetic barriers (Pb,Eu)Se},
  author = {Kazakov, Alexander and Volobuev, Valentine V. and Cho, Chang-Woo and Piot, Benjamin A. and Adamus, Zbigniew and Wojciechowski, Tomasz and Wojtowicz, Tomasz and Springholz, Gunther and Dietl, Tomasz},
  journal = {Phys. Rev. B},
  volume = {111},
  issue = {24},
  pages = {245419},
  numpages = {14},
  year = {2025},
  month = {Jun},
  publisher = {American Physical Society},
  doi = {10.1103/zb83-hf6p},
  url = {https://link.aps.org/doi/10.1103/zb83-hf6p}
}

@misc{Molly2000,
  title = {Veeco - {M}olly 2000},
  howpublished = {\url{https://www.veeco.com/products/growth-control-scheduling-software-for-mbe-systems/}},
  urldate = {2026-07-27}
}

@misc{GENxplor,
  title = {Veeco {GEN}xplor {R}\&{D} {MBE} {S}ystem},
  howpublished = {\url{https://www.veeco.com/products/genxplor-rd-mbe-system/}},
  urldate = {2026-07-27}
}

@article{walieh_2025,
    author = {Khaireh-Walieh, Abdourahman and Arnoult, Alexandre and Plissard, S{\'e}bastien and Wiecha, Peter R.},
    title = {Data-Driven Azimuthal RHEED Construction for In Situ Crystal Growth Characterization},
    journal = {Crystal Growth \& Design},
    volume = {25},
    number = {18},
    pages = {7438-7445},
    year = {2025},
    doi = {10.1021/acs.cgd.5c00368}
}

@article{Vasudevan_2014,
    author = {Vasudevan, Rama K. and Tselev, Alexander and Baddorf, Arthur P. and Kalinin, Sergei V.},
    title = {Big-Data Reflection High Energy Electron Diffraction Analysis for Understanding Epitaxial Film Growth Processes},
    journal = {ACS Nano},
    volume = {8},
    number = {10},
    pages = {10899-10908},
    year = {2014},
    doi = {10.1021/nn504730n}
}

@article{Muetzel_2026,
    author = {Muetzel, N. and Luu, V. and Bey, S. and Abdul Karim, M. and Yoshimura, K. and Liu, X. and Gebran, M. and Assaf, B. A.},
    title = {RHEED pattern classification by a convolutional neural network for the growth of chalcogenide thin films and nanostructures},
    journal = {AIP Advances},
    volume = {16},
    number = {3},
    pages = {035317},
    year = {2026},
    doi = {10.1063/5.0308043}
}

@article{Yoshinari_2026,
    author = {Yoshinari, A. and Ando, Y. and Matsumura, T. and Kotsugi, M. and Hasegawa, S. and Hobara, R. and Nagamura, N.},
    title = {Luminance histogram analysis for RHEED images using peak fitting approach with machine learning technique},
    journal = {Science and Technology of Advanced Materials: Methods},
    volume = {6},
    number = {1},
    pages = {2666924},
    year = {2026},
    doi = {10.1080/27660400.2026.2666924}
}

@article{Guo_2025,
    author = {Guo, Y. and Meisenheimer, P. and Qin, S. and Zhang, X. and Goddy, J. and Ramesh, R. and Martin, L. W. and Agar, J.},
    title = {Predicting Pulsed-Laser Deposition {SrTiO$_3$} Homoepitaxy Growth Dynamics Using High-Speed Reflection High-Energy Electron Diffraction},
    journal = {ACS Applied Materials \& Interfaces},
    volume = {17},
    number = {16},
    pages = {24485-24493},
    year = {2025},
    doi = {10.1021/acsami.4c12655}
}

@article{Harris_2024,
    author = {Harris, S. B. and Biswas, A. and Yun, S. J. and Roccapriore, K. M. and Rouleau, C. M. and Puretzky, A. A. and Vasudevan, R. K. and Geohegan, D. B. and Xiao, K.},
    title = {Autonomous Synthesis of Thin Film Materials with Pulsed Laser Deposition Enabled by In Situ Spectroscopy and Automation},
    journal = {Small Methods},
    volume = {8},
    number = {9},
    pages = {2301763},
    year = {2024},
    doi = {10.1002/smtd.202301763}
}

@article{Zheng_2025,
    author = {Zheng, Y. B. and Blake, C. and Mravac, L. and Zhang, F. and Chen, Y. and Yang, S.},
    title = {A self-driving physical vapor deposition system making sample-specific decisions on the fly},
    journal = {npj Computational Materials},
    volume = {11},
    pages = {327},
    year = {2025},
    doi = {10.1038/s41524-025-01805-0}
}

@article{Harris_2024_plume,
    author = {Harris, S. B. and Rouleau, C. M. and Xiao, K. and Vasudevan, R. K.},
    title = {Deep learning with plasma plume image sequences for anomaly detection and prediction of growth kinetics during pulsed laser deposition},
    journal = {npj Computational Materials},
    volume = {10},
    pages = {105},
    year = {2024},
    doi = {10.1038/s41524-024-01275-w}
}

@article{Daniluk_2025,
    author = {Daniluk, Andrzej and Daniluk, Bart{\l}omiej and W{\'o}jcik, Grzegorz M.},
    title = {A Python code for simulations of RHEED intensity oscillations within the one-dimensional dynamical approximation},
    journal = {Computer Physics Communications},
    volume = {308},
    pages = {109467},
    year = {2025},
    doi = {10.1016/j.cpc.2024.109467}
}

@article{Daniluk_2026,
    author = {Daniluk, Bart{\l}omiej and Daniluk, Andrzej and W{\'o}jcik, Grzegorz M.},
    title = {A Python package for simulations of RHEED intensity oscillations within the kinematical approximation},
    journal = {Computer Physics Communications},
    volume = {325},
    pages = {110186},
    year = {2026},
    doi = {10.1016/j.cpc.2026.110186}
}

@article{Kaufmann_2020,
    author = {Kaufmann, Kevin and Zhu, Chaoyi and Rosengarten, Alexander S. and Maryanovsky, Daniel and Harrington, Tyler J. and Marin, Eduardo and Vecchio, Kenneth S.},
    title = {Crystal symmetry determination in electron diffraction using machine learning},
    journal = {Science},
    volume = {367},
    number = {6477},
    pages = {564-568},
    year = {2020},
    doi = {10.1126/science.aay3062}
}

@article{Vecsei_2019,
    author = {Vecsei, Pascal M. and Choo, Kenny and Chang, Johan and Neupert, Titus},
    title = {Neural network based classification of crystal symmetries from x-ray diffraction patterns},
    journal = {Physical Review B},
    volume = {99},
    pages = {245120},
    year = {2019},
    doi = {10.1103/PhysRevB.99.245120}
}

@article{Ziatdinov_2017,
    author = {Ziatdinov, Maxim and Dyck, Ondrej and Maksov, Artem and Li, Xufan and Sang, Xiahan and Xiao, Kai and Unocic, Raymond R. and Vasudevan, Rama and Jesse, Stephen and Kalinin, Sergei V.},
    title = {Deep Learning of Atomically Resolved Scanning Transmission Electron Microscopy Images: Chemical Identification and Tracking Local Transformations},
    journal = {ACS Nano},
    volume = {11},
    number = {12},
    pages = {12742-12752},
    year = {2017},
    doi = {10.1021/acsnano.7b07504}
}

@article{Price_2024,
    author = {Price, Christopher C. and Li, Yansong and Zhou, Guanyu and Younas, Rehan and Zeng, Spencer S. and Scanlon, Tim H. and Munro, Jason M. and Hinkle, Christopher L.},
    title = {Predicting and Accelerating Nanomaterial Synthesis Using Machine Learning Featurization},
    journal = {Nano Letters},
    volume = {24},
    number = {46},
    pages = {14862-14867},
    year = {2024},
    doi = {10.1021/acs.nanolett.4c04500}
}

@article{Stuckner_2022,
    author = {Stuckner, Joshua and Harder, Bryan and Smith, Timothy M.},
    title = {Microstructure segmentation with deep learning encoders pre-trained on a large microscopy dataset},
    journal = {npj Computational Materials},
    volume = {8},
    pages = {200},
    year = {2022},
    doi = {10.1038/s41524-022-00878-5}
}

@article{Petkovic_2025,
    author = {Petkovic, M. and Vieira, L. and Dropka, N.},
    title = {Machine learning in crystal growth: A review of methods, data, and applications},
    journal = {Progress in Crystal Growth and Characterization of Materials},
    volume = {71},
    pages = {100689},
    year = {2025},
    doi = {10.1016/j.pcrysgrow.2025.100689}
}

@article{Bansal_2024,
    author = {Bansal, S. and Jain, A. and Kumar, S. and Kumar, A. and Kumar, P. R. and Prakash, K. and Soliman, M. S. and Islam, M. S. and Islam, M. T.},
    title = {Optoelectronic performance prediction of HgCdTe homojunction photodetector in long wave infrared spectral region using traditional simulations and machine learning models},
    journal = {Scientific Reports},
    volume = {14},
    pages = {28230},
    year = {2024},
    doi = {10.1038/s41598-024-79727-y}
}

@article{Wang_2020,
    author = {Wang, Anthony Yu-Tung and Murdock, Ryan J. and Kauwe, Steven K. and Oliynyk, Anton O. and Gurlo, Aleksander and Brgoch, Jakoah and Persson, Kristin A. and Sparks, Taylor D.},
    title = {Machine Learning for Materials Scientists: An Introductory Guide toward Best Practices},
    journal = {Chemistry of Materials},
    volume = {32},
    number = {12},
    pages = {4954-4965},
    year = {2020},
    doi = {10.1021/acs.chemmater.0c01907}
}

@misc{supp_info,
  title = {Supporting Information},
  author = {Turowski, Bartłomiej and Meixner, Jakub J. and Dziewiątkowska, Róża and Zaleszczyk, Wojciech and Wojciechowski, Tomasz and Volobuev, Valentine V. and Wysokiński, Marcin M. and Wojtowicz, Tomasz},
  url = {[link]},

}

\end{document}


\maketitle

\section*{Keywords}

Molecular Beam Epitaxy, RHEED, Computer Vision, Machine Learning, Real-Time Feedback, CdTe substrate

\pagebreak

\section*{Systematic features}

\begin{figure}[h]
    \centering
    \includegraphics[width=1\textwidth]{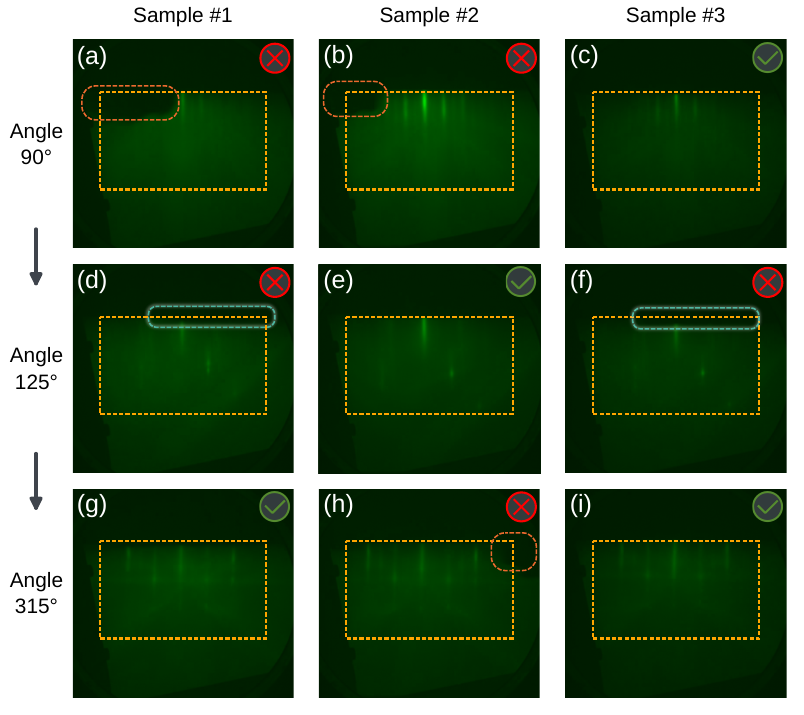}
    \caption{Set of three frames for three samples showing systematic features causing frames to be rejected ({\color{red} \faTimesCircleO}) before images processing pipeline described in section \textbf{III B}. \textbf{(a)} - \textbf{(c)} frames for angle $90^\circ$. \textbf{(d)} - \textbf{(f)} frames for angle $125^\circ$. \textbf{(g)} - \textbf{(i)} frames for angle $315^\circ$. \textbf{(a), (b), (h)} Three--handle molyblock holder shadow marked in {\color{magtop_orange}dashed orange} area. \textbf{(d), (f)} shadow of material flakes accumulating on the heater area marked in {\color{main_green}dashed green} area. \textbf{(c), (e), (g), (i)} frames that have no systematic issues and can be used in preprocessing pipeline ({\color{accept_green}\faCheckCircleO}).}
    \label{Sfgr:deleted_frames}
\end{figure}

In order to reduce risk of the neural network (NN) short-cutting the reflection high-energy electron diffraction (RHEED) pattern learning by correlating periodically occurring systematic features, shown in Figure \ref{Sfgr:deleted_frames}, frames with these features were removed in first step of data processing, before automated preprocessing pipeline.

\pagebreak

\section*{Dataset}
\label{supces:dataset}

List of all structures in the dataset with the number of rotations per structure and the number of frames used in NN training.

\begin{table}[htbp]
  \centering
  \renewcommand{\arraystretch}{1.4}
  \caption{RHEED single-rotation dataset: number of single rotation frame sets per structure with number of total frames per structure in the dataset and growth temperature $T_\mathrm{s}$. Structures were grown in October 2024 -- January 2025 time frame.}
  \label{tab:rheed_dataset}
  \begin{tabular}{l c r c l}
    \hline
    Structure code & Rotations & Images & $T_\mathrm{s}$ (\textcelsius) & Material/Substrate \\
    \hline
    V100224A & 8 & 7\,055 & 405 & CdTe/GaAs \\
    V100324A & 8 & 7\,102 & 412 & CdTe/GaAs \\
    V100424A & 8 & 7\,049 & 413 & CdTe/GaAs \\
    V100924A & 8 & 7\,154 & 405 & CdTe/GaAs \\
    V101024A & 8 & 7\,202 & 412 & CdTe/GaAs \\
    V101124A & 7 & 6\,234 & 420 & CdTe/GaAs \\
    V112624A & 7 & 6\,241 & 415 & CdTe:In/GaAs \\
    V112724A & 6 & 5\,239 & 420 & CdTe/GaAs \\
    V112824A & 7 & 6\,064 & 420 & CdTe/GaAs \\
    V120324A & 8 & 7\,058 & 410 & CdTe/GaAs \\
    V120424A & 7 & 6\,074 & 410 & CdTe/GaAs \\
    V120524A & 7 & 5\,970 & 410 & CdTe:In/GaAs \\
    V121724A & 7 & 6\,213 & 380 & CdTe/GaAs \\
    V011425A & 8 & 7\,231 & 410 & CdTe/GaAs \\
    V011525A & 7 & 6\,325 & 410 & CdTe/GaAs \\
    V011625A & 8 & 7\,286 & 410/350$^{\dagger}$ & CdTe:In/GaAs \\
    V012225B & 8 & 7\,280 & 398 & CdTe/GaAs \\
    V012325A & 8 & 6\,864 & 410 & CdTe/GaAs \\
    V012325B & 8 & 6\,708 & 400 & CdTe/GaAs \\
    V012925A & 8 & 7\,141 & 400 & CdTe/GaAs \\
    \hline
    \textbf{Total} & \textbf{151} & \textbf{133\,490} & -- & 20 structures \\
    \hline
  \end{tabular}
  \par\vspace{2pt}
  {\footnotesize $^{\dagger}$ Two-step growth: substrate held at the two listed temperatures in sequence.}
\end{table}

\pagebreak

\section*{G channel data extraction}

\begin{figure}[h]
    \centering
    \includegraphics[width=1\textwidth]{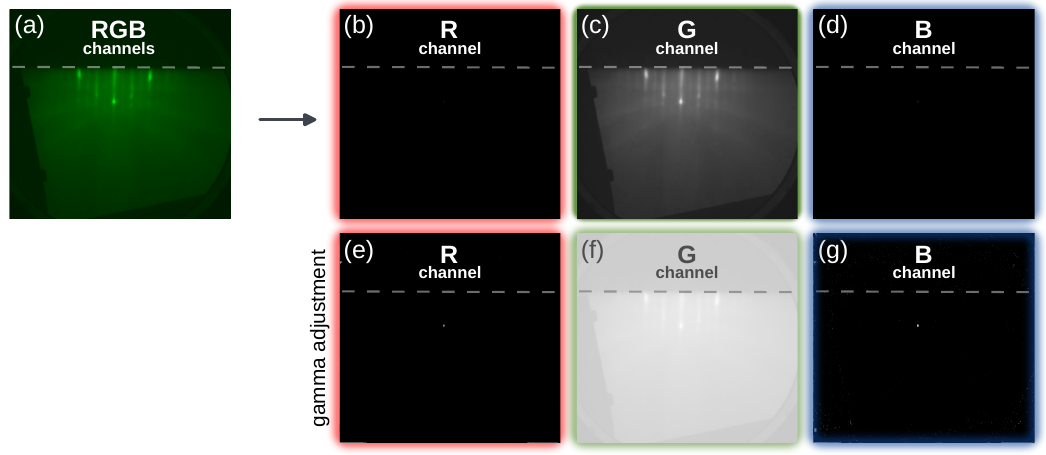}
    \caption{Green (G) channel extraction from raw three-channel data (RGB). \textbf{(a)} raw RHEED frame in RGB. \textbf{(b)-(d)} R, G, B extracted frames. \textbf{(e)-(g)} frames corresponding to (b) - (d) with gamma adjustment to make the R and B channels signal more visible.}
    \label{fgr:RGB-channels-split}
\end{figure}

97\% of signal is present in the G channel of the recorded RHEED videos, as illustrated in Figure \ref{fgr:RGB-channels-split}. Because of that fact, only the G channel is extracted for use in model training.

\section*{RHEED videos inference}

The proposed solution can be used both in real time during the molecular beam epitaxy (MBE) growth process and post factum, on recorded RHEED videos. Movies V112122A-inference\_2RPM.mp4 and V042426A-inference\_1RPM.mp4 (where RPM stands for Rotations Per Minute) illustrate the former case: the RHEED videos were played back with the VLC media player, and the trained ResNet2D model was used for inference. Real angle is shown in bottom-left corner of the video. The $0^\circ$ angle direction was determined by experienced RHEED operator and corresponds to $0^\circ$ angle direction in training data set.

A transparent frame-capturing overlay with an aspect ratio of 24:14 (width:height) was placed over the RHEED video. This aspect ratio matches that of the model input resolution of $96\times56$, so the captured region of interest (ROI) is resized to the model input without geometric distortion. From each captured frame the green channel is extracted, scaled to the $[0,1]$ range, resized with antialiased bicubic interpolation, and standardized using the global mean and standard deviation of the training set. The resulting single-channel image is then replicated across the R, G, and B channels and passed to the trained model to ensure architecture compatibility.

The NN was trained on the $[0^\circ, 180^\circ)$ range, with angles folded modulo $180^\circ$. As a consequence of this implementation decision, the raw inference output is capped at $180^\circ$ and displayed as the "Raw angle" readout. The "Kinematic" field shows the angle calculated from the intercept obtained with the postprocessing procedure described in \textbf{Section~V.~A.}

For completeness, we note that these videos were recorded two years before (V112122A, 21~November~2022) and more than a year after (V042426A, 24~April~2026) the videos used for model training (October 2024 -- January 2025).

Both videos used for inference verification were recorded at 2~RPM. For inference recording (V042426A-inference\_1RPM.mp4 and V112122A-inference\_2RPM.mp4), the original V042426A video was played back at half speed, emulating the 1~RPM rotation used during the growth procedure. The original video was recorded with the data-acquisition pipeline developed after the \textbf{Dataset} had been collected; the original V112122A video was recorded earlier, hence its lower average frame rate of 12.5 FPS (vs 18 FPS of the videos in \textbf{Dataset}). In both cases the intercept - equal to minus the angular distance remaining to the $0^\circ$ direction at the moment the inference is started - is recovered well within the error tolerance:
\begin{itemize}
    \item V042426A-inference\_1RPM.mp4 - inference started at Real angle
    $314^\circ$, expected intercept $-(360^\circ-314^\circ) = -46^\circ$,
    calculated intercept $-46.62^\circ$;
    \item V112122A-inference\_2RPM.mp4 - inference started at Real angle
    $340^\circ$, expected intercept $-(360^\circ-340^\circ) = -20^\circ$,
    calculated intercept $-20.02^\circ$.
\end{itemize}
Black padding was added to both videos so that the RHEED patterns appear at the same scale, making their comparison more convenient; the padding lies outside the captured ROI and does not enter the inference.

Both inference videos were recorded at 2RPM but for inference the V042426A-inference\_1RPM.mp4 is played at half speed. This video was recorded following data gathering pipeline developed after gathering the \textbf{Dataset}. V112122A-inference\_2RPM.mp4 was recorded earlier, so average frames per second (FPS) is lower: 12.5. Nevertheless, in both cases  the intercept is calculated extremely well: 
\begin{itemize}
    \item V042426A-inference\_1RPM.mp4 - start button is pressed at Real angle 314 deg, intercept is calulated as $-46.62$ deg. (360 - 314 = 46)
    \item V112122A-inference\_2RPM.mp4 - start button is pressed at Real angle 340 deg, intercept is calulated as $-20.02$ deg. (360 - 340 = 40)
\end{itemize}

Black padding is added for videos to ensure that RHEED pattern is of the same scale on both of them to make the patterns comparisons more convenient.

\section*{ROI placement robustness}

On-the-fly augmentation was introduced during training to emulate the operator-dependent variability in how the ROI is placed over the RHEED video stream, making the raw NN inference more robust to ROI placement. Video V042426A-ROI\_robustness\_NO\_postprocessing.mp4 illustrates this robustness on the eight main crystallographic directions (i.e., directions aligned with the crystallographic axes of the grown material) of a CdTe~(100) buffer. In contrast to the inference videos described above, each of the eight segments shows a single static RHEED frame, over which only the capturing ROI is moved. The angle readout displayed for every ROI position is the raw NN inference of the currently captured region, with no postprocessing involved. The true azimuth of each static frame - the inference target - is displayed as the ``Real angle'' value and serves as the reference for the inferred angle. Since the underlying frame does not change, any variation of the inferred angle directly reflects the sensitivity of the raw inference to ROI placement.

As noted above, the NN was trained on the $[0^\circ, 180^\circ)$ range with angles folded modulo $180^\circ$, so the raw inference output is capped at $180^\circ$. As a consequence, for directions near $0^\circ$ the raw readout may appear just below $180^\circ$ (cf. segment (a)), since $0^\circ$ and $180^\circ$ are equivalent modulo $180^\circ$.
Video (e) shows the Real angle readout of $186^\circ$ because the three-handle molyblock holder (shadow visible in the top-left corner of the ROI) was obstructing the pattern at $180^\circ$.